\documentclass[aps,prd,preprint,tightenlines,groupedaddress,showpacs]{revtex4}
\usepackage{epsf,epsfig,graphics,graphicx}
\usepackage{verbatim,color,ulem}
\begin{document}
\title{Photon production from
non-equilibrium \\disoriented chiral condensates in a longitudinal
expansion : A theoretic framework}
\author{ Da-Shin Lee}
\email{ dslee@mail.ndhu.edu.tw} \thanks{ corresponding  author}
\affiliation{ Department of Physics, National Dong Hwa University,
Hua-Lien, Taiwan 97401, R.O.C. }
\author{Yeo-Yie Charng}
\email{charng@phys.sinica.edu.tw}
 \author{Kin-Wang Ng}
  \email{nkw@phys.sinica.edu.tw}
  \affiliation{ Institute of Physics,
Academia Sinica, Taipei, Taiwan 11529, R.O.C. }

\date{\today}

\begin{abstract}
A theoretical framework is developed for treating the quantization
of the photons in a spacetime with a longitudinal expansion. This
can be used to study the production of the photons through the
non-equilibrium relaxation of a disoriented chiral condensate
presumably formed in the expanding hot central region in
ultra-relativistic heavy-ion collisions. These photons can be a
signature of the formation of disoriented chiral condensates in the
direct photon measurements of heavy-ion collisions.
\end{abstract}

\maketitle

Disoriented Chiral Condensates (DCCs), the correlated space-time
regions where the chiral order parameter of QCD is chirally rotated
from its  orientation  of the true vacuum state in the isospin
space, might be formed in high energy hadronic or nuclear
collisions. The search of such a DCC would provide a probe to the
understanding of the  chiral structure of the QCD vacuum and/or the
chiral phase transition of strong interactions at high temperatures.
Various  experimental searches of  this phenomenon include cosmic
ray experiments, the MiniMAX experiment in proton-antiproton
collisions at the Fermilab Tevatron, or event-by-event analysis of
Pb-Pb collisions by the WA98 and NA49 collaborations at the CERN
SPS. Although no clear experimental evidence for DCC formation has
been reported so far, the search of this phenomenon is still part of
the missions in the heavy-ions physics program in present or planned
experiments at RHIC and LHC~\cite{ale,moh}.

In relativistic heavy ion collisions, the large energy deposit in
the central rapidity region leaves behind a hot and dense plasma at
a temperature above $200 ~{\rm MeV}$ where the chiral symmetry is
restored. Then, the plasma cools down in a rapid hydrodynamic
expansion accompanied by the chiral phase transition. The subsequent
out of equilibrium evolution triggers an exponential growth of
long-wavelength fluctuations via the spinoidal
instabilities~\cite{wilczek,cooper,boydcc}, resulting in the
formation of disoriented chiral condensates (DCCs) (see also
Refs.~\cite{bjorken1,dcc}). The decay of such DCCs to the true QCD
vacuum is expected to radiate copiously soft pions that could be a
potential observational signature of the chiral phase transition.
However, the emitted pions will strongly interact with the
background of other hadronic matter so that the hadronic signal may
be severely masked. In contrast, electromagnetic probe such as
photon and lepton with longer mean free path in the medium serves as
a good candidate and can reveal more detailed non-equilibrium
information on the DCCs with minimal
distortion~\cite{wang,boy1,boy2}.

It has been proposed that anomalous radiation of low-momentum photon
and/or low-mass dileptons can be produced from a DCC. In particular,
in Ref.~\cite{boy2}, Boyanovsky {\it et al.} have extensively
studied the photon production from the DCC with an nonvanishing
expectation value of the neutral pion through the ${\rm U}_{\rm A}
(1)$ anomalous vertex. Later, the authors in Ref.~\cite{leeng} have
taken into account another dominant contribution that also involves
the dynamics of $\pi^{0}$ due to the decay of the vector meson
through the electromagnetic vertex. They have found that for large
initial amplitudes of the $\pi^0$ mean field the photon production
is enhanced by parametric amplification, resulting in a distinct
energy distribution of the produced photons. However, the above two
works have ignored the hydrodynamical expansion and adopted the
simple ``quench'' approximation for the chiral phase transition.
This quenched phase transition has been widely used in the study of
non-equilibrium phenomena of DCCs~\cite{boydcc,wang,boy1,boy2}.
Based upon the Bjorken's scenario, in ultra-relativistic heavy-ion
collisions, an approximate Lorentz boost invariant particle flow
along the longitudinal direction might be created in the central
region~\cite{bjor}. At late times following the heavy nuclei
collisions, a transverse flow can be generated due to the
multi-scattering between the produced particles, as such the
expansion becomes three dimensional~\cite{soll}. In
Ref.~\cite{charng}, the authors have considered the effect of the
hydrodynamical expansion of the plasma to the production of photons
from the non-equilibrium relaxation of a DCC. It is found that the
expansion smoothes out the resonances in the process of parametric
amplification such that the non-equilibrium photons are dominant to
the thermal photons over the range $0.2-2$ GeV. This work is to
assume a spherical boost invariant hydrodynamical flow which for
small values of the rapidity is conformally flat, thus greatly
simplifying the treatment of the photon field in the expanding
spacetime. In this work, we attempt to consider the nonequilibrium
photon production from a DCC in a more realistic longitudinal boost
invariant hydrodynamical expansion,  which certainly requires a
quantization scheme for electromagnetic fields under this
anisotropic expanding background.

The study of the quantized electromagnetic fields interacting with
the gravitation fields or propagating in the curved spacetime is of
interest in many different aspects, one of which is to consider the
photon production in the early universe with its possible effect on
the initial anisotropies~\cite{zel}. Thus  a consistent quantization
scheme of electromagnetic theory under an anisotropic expanding
background is required. In Ref.~\cite{lot}, quantization of
electromagnetic fields in a diagonal Bianchi type I metric is
proposed using the Fourier mode expansion of the fields developed in
Ref.~\cite{sag}. Applying the WKB particle concept in asymptotic in
and out regions of spacetime where the background expansion is
adiabatically slow gives the normalization of the mode functions.
Then particle production of the free photons is considered by
performing the standard Bogoliubov transformation between
annihilation and creation operators in in and out regions. Here we
instead try to provide an alternative quantization scheme directly
on electromagnetic potentials. An application of this quantization
method to interactive photon fields  is straightforward, in
particular when some other charged matter fields are also present
with the interaction term depending upon electromagnetic potentials.

After introducing our model and summarizing the dynamical equations
for describing the evolution of the expectation value of the fields
within the DCCs, in this {\it Letter} we will propose the
quantization method on quantizing electromagnetic potentials in an
anisotropic expanding background and discuss its application to the
interactive photons.  The formalism for considering photon
production will be developed. Summary and discussions on our future
investigation will be presented at the end.

\vskip 0.5cm

The longitudinal expansion, say along the $z$ axis, of the boost
invariant hydrodynamical flow can be described by  the proper time
$\tau$ and the space-time rapidity $\eta$, defined as
\begin{equation}
\tau\equiv (t^2-z^2)^{1\over2},\quad
\eta\equiv {1\over2}\ln\left(\frac{t+z}{t-z}\right),
\end{equation}
where $(t,{\bf x})=(t,{\bf x}_\bot,z)=(t,x,y,z)$ are the coordinates
in the laboratory,
\begin{equation}
t=\tau\cosh\eta,\quad z=\tau\sinh\eta.
\label{transf}
\end{equation}
The ranges of these coordinates are set to be  $0 \le \tau < \infty$
and $0 \le \eta < \infty$, restricted  to the forward light cone.
The Minkowski line element is then given by
\begin{eqnarray}
ds^2&=&dt^2-d{\bf x}^2 \nonumber \\
    &=&d\tau^2-d{\bf x}_\bot^2-\tau^2 d\eta^2.
\label{kasner}
\end{eqnarray}
Eq.~(\ref{kasner}) is the Kasner metric that serves as the
background comoving frame under which non-equilibrium photons are
to be produced from a DCC domain.

The dynamics of the DCC as well as the photon production from a DCC
in a generally expanding background can be described by the
phenomenological action given by
\begin{equation}
S=\int d^4x {\sqrt g} \left( L_\sigma+ L_A + L_{\pi^0 A}\right) \, ,
\label{action}
\end{equation}
where
\begin{eqnarray}
L_\sigma &=& -{1\over2}g^{\mu\nu}\partial_\mu {\vec\Phi}\cdot \partial_\nu
             {\vec\Phi} + {1\over2} \frac{M_\sigma^2}{2}
             {\vec\Phi}\cdot {\vec\Phi}
             - \lambda \left({\vec\Phi}\cdot{\vec\Phi}\right)^2 + h\sigma, \\
L_A &=& -{1\over4} g^{\alpha\mu} g^{\beta\nu} F_{\alpha\beta} F_{\mu\nu}, \\
L_{\pi^0 A} &=& \frac{1}{\sqrt g}\frac{e^2}{32\pi^2}\frac{\pi^0}{f_\pi}
             \epsilon^{\alpha\beta\mu\nu} F_{\alpha\beta} F_{\mu\nu}
             +\frac{1}{(\sqrt g)^2} \frac{e^2\lambda_V^2}{8m_\pi^2 m_V^2}
             \epsilon^{\mu\nu\lambda\sigma}\epsilon^{\alpha\beta\gamma\delta}
             g_{\sigma\delta} \partial_\lambda \pi^0 \partial_\gamma \pi^0
             F_{\mu\nu}F_{\alpha\beta} \, .
\end{eqnarray}
${\vec\Phi}=(\sigma,\pi^0,{\vec\pi})$ is an $O(N+1)$ vector of
scalar fields with $\vec\pi=(\pi^1,\pi^2,...,\pi^{N-1})$
representing the $N-1$ pions.  The large-$N$ approximation will be
implemented later to account for the strong coupling effects among
the $\sigma $ and all other $\pi $ fields. The phenomenological
parameters in the effective Lagrangian above can be determined by
the low-energy pion physics. $F_{\mu\nu}=\partial_\mu A_\nu -
\partial_\nu A_\mu$ is the electromagnetic field tensor. The
dominant couplings between the photon and the neutron pion are given
by the anomalous $U_A (1)$ interaction as well as  the decay of the
vector meson $V_{\mu}$.  $V_{\mu}$  is identified as the $\omega$
meson with its coupling $\lambda_{V}$ that is obtained from the $
\omega \rightarrow \pi^0 \gamma$ decay width~\cite{davi}. The
signature of the metric we choose is $(-+++)$ where $ g=-{\rm det} [
g_{\mu\nu}]= \tau^2$. In particular, since
$\epsilon^{\alpha\beta\mu\nu}$ is a tensor density of weight $-1$,
each of this pseudo tensor is multiplied  by the factor $1/{\sqrt
g}$ in a general coordinate system~\cite{wein}. The above effective
interactions on the photon have been studied by us and are obtained
from the perturbative theory without involving in-medium
modifications. The in-medium effects will enter only through the
non-equilibrium fluctuations for the pion fields due to their strong
couplings .

Within the DCC, we assume that the $\sigma$ and the $\pi^0$ acquire
the respective expectation values that both are a function of the
comoving time $\tau$ only in such a boost invariant background. Here
we first shift $\sigma$ and $\pi^0$ by their expectation values with
respect to an initial non-equilibrium states:
\begin{eqnarray}
&&\sigma({\bf x},\tau) = \phi(\tau) + \chi({\bf x},\tau),\quad\quad
  \langle\sigma({\bf x},\tau)\rangle = \phi (\tau), \\
&&\pi^0({\bf x},\tau) = \zeta(\tau) + \psi({\bf x},\tau),\quad\quad
  \langle\pi^0({\bf x},\tau)\rangle = \zeta(\tau),
\end{eqnarray}
with the tadpole conditions:
\begin{equation}
\langle\chi({\bf x},\tau)\rangle=0,\quad \langle\psi({\bf
x},\tau)\rangle=0,\quad \langle{\vec \pi}({\bf x},\tau)\rangle=0.
\label{tad}
\end{equation}
Using the Schwinger-Keldysh closed-time-path formulation of
non-equilibrium quantum field theory allows us to derive the
evolution equation of the non-equilibrium expectation values as
well as the correlation functions of quantum fields. This
technique has successfully been employed elsewhere within many
different contexts and we refer the readers to the literature for
details~\cite{boyneq}. To self-consistently incorporate quantum
fluctuation effects from the strong $\sigma-\pi$ interactions, we
then implement the large-$N$ approximation that provides a
nonperturbative resummation scheme below (Eqs.~(2.7)-(2.9) of
Ref.~\cite{boy2}). Using the tadpole conditions~(\ref{tad}), the
large-$N$ equations of motion for the mean fields that also
involve the perturbatively electromagnetic corrections  can be
obtained as follows:
\begin{eqnarray}
&& \left[ \partial^2_{\tau} +\frac{1}{\tau}
\partial_{\tau}-\frac{M_\sigma^2}{2} +4\lambda\phi^2(\tau)
+4\lambda\zeta^2(\tau) +4\lambda\langle{\vec \pi}^2\rangle(\tau)
 \right] \phi (\tau) -  h=0, \nonumber \\
&&\left[  \partial^2_{\tau} +\frac{1}{\tau}
\partial_{\tau}-\frac{M_\sigma^2}{2}+4\lambda\phi^2(\tau)
+4\lambda\zeta^2(\tau)
+4\lambda\langle{\vec \pi}^2\rangle(\tau)\right] \zeta (\tau) \nonumber\\
&&\quad\quad \quad\quad -\frac{e^2}{32\pi^2 f_\pi}
 \langle F {\tilde F} \rangle (\tau)
+\frac{e^2\lambda_V^2}{4m_\pi^2 m_V^2} \tau^{-1}
\partial_\tau \left[\tau^{-1} \dot{\zeta}\right] \epsilon^{\mu\nu 0\sigma}\epsilon^{\alpha\beta
0\delta} g_{\sigma\delta} \langle F_{\mu\nu}F_{\alpha\beta}\rangle
(\tau)
\nonumber\\
&& \quad\quad \quad\quad\quad\quad
\quad\quad\quad+\frac{e^2\lambda_V^2}{4m_\pi^2 m_V^2}
\tau^{-2}\dot{\zeta} \, \epsilon^{\mu\nu
0\sigma}\epsilon^{\alpha\beta \gamma\delta}g_{\sigma\delta} \langle
\partial_{\gamma} F_{\mu\nu}F_{\alpha\beta}\rangle (\tau)=0 \,,
\label{eommf}
\end{eqnarray}
where the dot means $d/d\tau$. The backreaction from the quantum
fluctuations of the pions can be expressed by their Fourier mode
functions $U_{ k}(\tau)$ defined as:
\begin{equation}
{\vec \pi}({\bf x},\tau) = \int \frac{d^3{ k}}{(2\pi)^{\frac{3}{2} }
} \frac{1}{\sqrt{2 \omega_{{\vec\pi} k}(\tau_0)}}  \left[\, {\vec
a}_{\bf k} \, U_{ k}(\tau) \, e^{i\bf k\cdot \bf x} + {\rm
h.c.}\right]\, ,
\end{equation}
where $\bf{k} \cdot \bf{x} = \bf{k}_\bot \bf{x}_\bot + k_\eta \eta $
and $d^3{\vec k}=d^2 k_\bot d k_\eta $. ${\vec a}_{\bf k}$ is a
destruction operator for the mode with frequency $\omega_{{\vec
\pi}\vec k}(\tau_0)$  to be determined from
solving the time independent gap equation given by  the initial
states. The mode equations for the pions can be read off from the
quadratic part of the effective Lagrangian as
\begin{equation}
\left[ \partial^2_{\tau} +\frac{1}{\tau}
\partial_{\tau}+ k^2-\frac{M_\sigma^2}{2} +4\lambda\phi^2(\tau)
+4\lambda\zeta^2(\tau) + 4\lambda\langle{\vec
\pi}^2\rangle(\tau)\right]U_{k} (\tau)=0,  \\
 \label{mepi}
\end{equation}
where $k^2 = g_{\mu \nu} k^\mu k^\nu = k_\bot^2 +k_\eta^2/\tau^2$.
The expectation values with respective to the initial states are
given by
\begin{equation}
\langle{\vec \pi}^2\rangle(\tau) = (N-1) \int^{\Lambda } \frac{d^3{
k}}{2(2\pi)^3\omega_{{\vec \pi}{k}}(\tau_0)} \left[ |U_{ k}(\tau)|^2
\right] \coth \left[ \frac{\omega_{{\vec \pi} k}(\tau_0)}{2 T_i}
\right]  \, , \label{pifluc}
\end{equation}
where $\langle{\vec \pi}^2\rangle(\tau)$ will be self-consistently
determined by the above equation with an initial temperature set to
$T_i$ presumably above the critical temperature of the chiral phase
transition. Thus, the corresponding the gap
equation for the initial adiabatic modes is
\begin{equation}
\omega^2_{{\vec \pi} k}(\tau_0)=k^2-\frac{M_\sigma^2}{2}
+4\lambda\phi^2(\tau_0) +4\lambda\zeta^2(\tau_0) +
4\lambda\langle{\vec \pi}^2\rangle(\tau_0) \, .
\end{equation}
It must be noticed that this is an effective field theory with an
ultraviolet momentum cutoff of the order of $\Lambda \approx m_{V}$.
Without considering the perturbative backreaction effects from the
photons, the above equations have been extensively studied in
Ref.~\cite{cooper}, aiming at finding the phase space of initial
conditions that may lead to instabilities on long wavelength modes
during the subsequent nonequilibrium chiral phase transition. If so,
then the DCCs may be formed, and thus produce an anomalous
transverse distribution of secondary pions when compared to a more
conventional boost invariant hydrodynamic flow in local thermal
equilibrium. Here these background solutions are of importance to
provide a nonequilibrium environment from which the photons are to
be produced.

The relevant Lagrangian densities for describing the dynamics of
photon production under the time dependent expectation value of the
neutral pion  are given by $L_A$ and $L_{\zeta A}$ in particular
where the pion field in $L_{\pi^0 A}$ is replaced by its mean value
$\zeta$. The underlying gauge symmetry embedded in electromagnetic
theory reflects the fact that not all gauge potential are of
physical relevance. Thus we can avoid potential gauge ambiguities by
eliminating all possible redundant degrees of freedom with a choice
of the gauge fixing.

The canonical momenta, $\Pi^\mu = ( \Pi^\tau, \Pi^i)$, conjugate to
the potential fields, $A_\mu = ( A_\tau, A_i)$, are defined by
\begin{eqnarray}
\Pi^{\tau} &=& \frac{\delta}{ \delta \dot{A}_{\tau} } \sqrt{g}
\left( L_{A} + L_{\zeta A} \right)=0 \, , \\
\Pi^i &=& \frac{\delta}{ \delta \dot{A}_{i} } \sqrt{g} \left( L_{A}
+ L_{\zeta A} \right) =\sqrt{g} \, (\,  g^{ij} F_{0 j} +
\frac{1}{\sqrt{g}} \frac{e^2}{ 8\pi^2 f_{\pi}} \, \zeta \,
\epsilon^{oijk} F_{jk} \, ) \, .
\end{eqnarray}
However, since $A_\tau $ is not a dynamical variable as its
conjugate momentum $\Pi^\tau$ vanishes, we can set $A_\tau =0$, a
temporal gauge. Then the further constraint equation can be
obtained by taking the variation of the Lagrangian density with
respect to $A_\tau $ that gives the Gauss law $
\partial_i \Pi^i=0$. Thus, the choice of the gauge fixing can be
such as
\begin{equation}
 A_\tau=0 \, ;\,\,\, \,\,\, g^{ij}
\partial_i \dot{A}_j =0 \, , \label{GF}
\end{equation}
the counterpart of the Coulomb gauge in a flat spacetime.  The gauge
conditions leave us two transverse vector potentials. It proves more
convenient to parametrize the polarization vectors in a longitudinal
expanding background as follows:
\begin{eqnarray}
{\bf{\varepsilon}}^{(1)}_{\theta  \, {\bf k} \, i }= \frac{1}{k
\tau} \left( \frac{k_x k_\eta}{k_\bot },\frac{ k_y k_\eta}{k_\bot},
-k_\bot \tau^2 \right) \, , \,\,\,\, \varepsilon^{(2)}_{\phi \,{\bf
k} \,i} =\left( -\frac{k_y}{k_\bot}, \frac{k_x}{k_\bot}, 0 \right)
\, , \,\,\,\, \varepsilon^{(3)}_{r \,{\bf k}\, i} =\frac{1}{k}
\left( k_x , k_y . k_\eta \right) \, .
\end{eqnarray}
They obey
\begin{equation}
\varepsilon^{(1)}_{{\bf -k}\, i} = \varepsilon^{(1)}_{{\bf k}\, i}
\,,\,\,\,\,\,\,\varepsilon^{(2)}_{{\bf - k}\, i} =-
\varepsilon^{(2)}_{{\bf k}\, i} \, , \,\,\,\,\,\,\,
 \varepsilon^{(a)\, \, l}_{\bf k} =\frac{1}{\sqrt{g}} \,
\epsilon^{lmn} \, \varepsilon^{(b)}_{{\bf k} \, m} \,
\varepsilon^{(c)}_{{\bf k} \, n } \, ,
\end{equation}
where $(a,b,c)=(1,2,3)$ and its cyclic order. Notice that
\begin{equation}
  \dot{\varepsilon}^{(1)}_{{\bf k} \,i} = \frac{k_\bot^2}{k^2
\tau} \left( \varepsilon^{(1)}_{{\bf k}\,i}-\frac{2 k_\eta}{k_\bot
\tau} \varepsilon^{(3)}_{{\bf k}\,i} \right)\, ,
\end{equation}
so the longitudinal expansion might lead to the change of the
direction of the polarization vector $\varepsilon^{(1)}_{{\bf
k}\,i}$, but keeps the other two polarization vectors
$\varepsilon^{(2)}_{{\bf k}\,i}$ and $\varepsilon^{(3)}_{{\bf
k}\,i}$ remaining their directions. Thus, the mode expansion of the
vector potentials must involve not only the state with polarization
 vector $\varepsilon^{(2)}_{ {\bf  k} \,i }$, but also the mixed
 state with the respective polarization vectors $\varepsilon^{1)}_{
{\bf k}\,i }$ and $\varepsilon^{(3)}_{ {\bf k} \, i }$, given by
\begin{equation}
A_i ({\bf{x}},\tau)= \int \frac{d^3 k}{( 2\pi)^{\frac{3}{2}}}
\frac{1}{\sqrt{2 k (\tau_0)}}  \left\{ \left[ b_{1 {\bf k}} \left(
V_{1 k}(\tau) \varepsilon^{(1)}_{{\bf{ k}} \,i}+ V_{3  k}(\tau)
\varepsilon^{(3)}_{ {\bf{ k}} \, i} \right)+ b_{2 \bf k} V_{2
k}(\tau) \varepsilon^{(2)}_{ {\bf{k}} \,i} \right]
 e^{i\bf k\cdot \bf x} + {\rm h.c.} \right\}\, .
\end{equation}
 The creation and annihilation operators satisfy the canonical
commutation relations. The mode function $ V_{3 \bf k}(\tau)$
certainly is not an independent function and its evolution depends
on the choice of the gauge that gives
\begin{equation}
\dot{V}_{3  k}(\tau)+ \frac{k^2_\eta}{ k^2 \tau^3} V_{3  k}(\tau)
=\frac{ 2 k_\bot k_\eta }{ k^2 \tau^2} V_{1  k}(\tau) \, .
\label{v3}
\end{equation}
It is then expected that  the physical degrees of freedom are those
two transverse modes with the polarization vectors obeying the
transversality condition,
\begin{equation}
     \sum_{a=1,2} \varepsilon^{(a)}_{{\bf k}\, i} \varepsilon^{(a)}_{{\bf k} \,j} = g_{ij} -\frac{k_i k_j}{ k^2} = P_{ij}({\bf k}) \,
     .
\end{equation}
It will be seen later that the equation (\ref{v3}) obtained from the
gauge fixing then becomes crucial in the sense that the expression
of the gauge invariant quantity of interest, for example electric
and magnetic fields,  have no $ V_{3  k}$ dependence.

Then the mode expansion of the conjugate momenta can be obtained
as
\begin{equation}
\Pi^i = E^i  + \frac{e^2}{ 4 \pi^2 f_{\pi}} \, \zeta \,  B^i \, ,
\end{equation}
where the electric and magnetic fields can be expressed  in terms of
two transverse degrees of freedom as follows:
\begin{eqnarray}
E^i (\bf{x},\tau) &=& \sqrt{g} g^{ij} F_{0j} \nonumber\\
&=& \int \frac{d^3 k}{( 2\pi)^{\frac{3}{2}}} \frac{1}{\sqrt{2 k
(\tau_0)}} \sqrt{g} \left\{ \left[ b_{1 \bf k} \left( \dot{V}_{1
k}(\tau) + \frac{k_\bot^2}{ k^2 \tau } V_{1 k}(\tau) \right)
\varepsilon^{(1)\,  i}_{\bf{ k} } + b_{2 \bf k} \dot{V}_{2 k}(\tau)
\varepsilon^{(2) \, i}_{ \bf{k} } \right]
 e^{i\bf k\cdot \bf x} + {\rm h.c.}\right\} \,  , \nonumber\\
\label{EField} \\
B^i (\bf{x},\tau) &=& \frac{1}{2}  \epsilon^{ijk} F_{jk} \,
\nonumber\\
&=& \int \frac{d^3 k}{( 2\pi)^{\frac{3}{2}}} \frac{1}{\sqrt{2 k
(\tau_0)}} \sqrt{g} \left\{ i  k \left[ b_{1 \bf k} {V}_{1 k}(\tau)
\varepsilon^{(2)\,  i}_{\bf{ k} } - b_{2 \bf k} V_{2 k}(\tau)
\varepsilon^{(1)\, i}_{ \bf{k} } \right]
 e^{i\bf k\cdot \bf x} + {\rm h.c.}\right\} \, , \label{BField}
\end{eqnarray}
so that the fields always point to the direction perpendicular to
the direction of their propagation ${\bf k}$.

The commutation relation between the vector potentials and their
conjugate momenta can be straightforwardly computed as follows:
\begin{eqnarray}
&& [\, \Pi^i_{\bf{k}} , A_{-{\bf k} j}\, ] = -i \hbar P^i_j ({\bf
k})\nonumber\\
&& \quad +\sqrt{g} \frac{1}{2 k(\tau_0)} \left\{ \left[ \left(
\dot{V}_{1 k}(\tau) + \frac{k_\bot^2}{ k^2 \tau } V_{1  k}(\tau)
\right) \, \varepsilon^{(1)\, \, i}_{{\bf k} } + \frac{ e^2}{ 4
\pi^2 f_{\pi}} \, \zeta \, i\, k  \, V_{1
 k}(\tau)  \, \varepsilon^{(2)\, i}_{\bf{ k} } \right]  V^*_{3
k}(\tau) \,\varepsilon^{(3)}_{ {\bf k} \, j}
\right. \nonumber\\
&&\quad\quad \quad\quad\quad  -\left. \left[ \left( \dot{V}^*_{1
k}(\tau) + \frac{k_\bot^2}{ k^2 \tau } V^*_{1 k}(\tau) \right)\,
\varepsilon^{(1)\, \, i}_{{\bf k} } + \frac{  e^2}{ 4 \pi^2 f_{\pi}}
\, \zeta \, i\,k V^*_{1 k}(\tau)\, \varepsilon^{(2)\, i}_{{\bf k } }
\right] V_{3 k}(\tau) \, \varepsilon^{(3)}_{ {\bf k} \, j} \right\}
\, ,\label{CR}
\end{eqnarray}
where the Wranskian conditions of the mode functions $ V_{1
k}(\tau)$ and $ V_{2  k}(\tau) $ and their time derivatives $
\dot{V}_{1 k}(\tau)$ and $ \dot{V}_{2 k}(\tau) $ are given by
\begin{eqnarray}
 \sqrt{g} \frac{1}{2 k(\tau_0)} \left[ \dot{V}_{1  k}(\tau) V^*_{1
 k}(\tau) - \dot{V}^*_{1  k}(\tau) V_{1  k}(\tau) \right] &=& -i \hbar \, , \label{wsv1} \\
\sqrt{g} \frac{1}{2 k (\tau_0)} \left[ \dot{V}_{2  k}(\tau) V^*_{2
k}(\tau) - \dot{V}^*_{2  k}(\tau) V_{2  k}(\tau) \right] &=& -i
\hbar \, . \label{wsv2}
\end{eqnarray}
The Wranskian equations are  to fix the normalization conditions of
the mode functions. The first term of the commutation relation
above~(\ref{CR}) is expected whereas the additional terms with the
$V_{3 k}(\tau) $ dependence are present to satisfy the gauge choice,
that is, $[\, \Pi^i_{\bf{k}} \,,\, k^j \dot{A}_{-{\bf k} j}\, ] =0$
obtained by means of its equation (\ref{v3}). Notice that since  we
are interested in the dynamics of photon production from a DCC that
does not necessarily involve the intermediate photon states, it will
be seen later that the expression of the number density of the
produced photons depends only on the dynamics of the mode functions
respectively, $V_{1 k}(\tau) $ and $V_{2 k}(\tau) $.

Although the quantization scheme can be formulated in the way that
all redundant degrees of freedom can be eliminated consistently by
choosing a proper gauge fixing, there might exist a subtlety on
finding a suitable mode expansion to further diagonalize
Lagrangian density of the interactive photon fields. This subtlety
will become more clear later  as the above quantization scheme is
applied to free photons first and
 to interactive photons of interest later. Besides, we will
mention the known alternative quantization scheme in
Ref.~\cite{lot} that is proposed for investigating the dynamics of
photon production under the more general time dependent
anisotropic spacetime, and comment on the advantage of our
formulation.

The Lagrangian density for free electromagnetic
fields $L_{A}$  can be obtained by imposing the gauge conditions~(\ref{GF})  as
\begin{equation}
L_{A}= \frac{1}{2} \dot{A}_{T \, i}
\dot{A}_{T}^i -\frac{1}{2} \partial_l  A_{T\, i} \partial^l  A_{T}^i
 \, ,
\end{equation}
where the equations of motion for the Fourier modes of the vector
potentials ${\bf A}_{T {\bf k}}$ are obtained as follows:
\begin{equation}
\partial^2_{\tau}  A_{T {\bf k}}^i (\tau)+\frac{1}{\tau}
\partial_{\tau}  A_{T {\bf k}}^i (\tau) + k^2  A_{T{\bf k}}^i (\tau)=0 \, .
\end{equation}
The mode functions for the linear polarization states are thus
\begin{eqnarray}
\ddot{V}^0_{1 k} (\tau) +\frac{1}{\tau} \dot{V}^0_{1 k} (\tau) +
\left( \frac{2 k^2_{\bot}}{ k^2 \tau^2} - \frac{3 k^4_{\bot}}{ k^4
\tau^2}+
k^2 \right) V^0_{1 k} (\tau) &=& 0 \, , \label{emv1} \\
 \ddot{V}^0_{2 k}
(\tau) +\frac{1}{\tau} \dot{V}^0_{2 k} (\tau) + k^2 V^0_{2 k} (\tau)
&=& 0 \, , \label{emv2}
\end{eqnarray}
where the Wranskian conditions~(\ref{wsv1}) and (\ref{wsv2}) are
satisfied. In Ref.~\cite{lot}, the quantization scheme is based upon
the method developed in Ref.~\cite{sag} where the mode functions $
S_{(\pm) k}$ to be defined later are used in the Fourier expansion
of the electric and magnetic fields.  The mode functions $ S_{(\pm)
k}$ can be related to the mode functions $V^0_{1k}$ and $V^0_{2k}$
as
\begin{equation}
S_{(\pm)k} = \frac{1}{2} \left[ \frac{k_\bot^2}{k^3 \tau} V^0_{1k}
+\frac{1}{k} \dot{V}^0_{1k} \mp V^0_{2k} \right] \,  . \label{SpmkV}
\end{equation}
Then, using Eqs.~(\ref{emv1}) and (\ref{emv2}), $\dot{S}_{(\pm)\, k}
$ can also be expressed as
\begin{equation}
\dot{S}_{(\pm)\, k} = -\frac{1}{2} \left[ k \, V^0_{1k} \pm
\dot{V}^0_{2k} \right] \, , \label{SdotpmkV}
\end{equation}
and the evolution equations of $ S_{(\pm) k}$ are found remarkably
simple, that is,
\begin{equation}
\ddot{S}_{(\pm)k}(\tau) +\frac{1}{\tau} \dot{S}_{(\pm)k}(\tau) + k^2
{S}_{(\pm)k}(\tau)=0 \, , \label{Spmk}
\end{equation}
obtained in Ref.~\cite{sag}. The corresponding Wranskian conditions
of the mode functions $ S_{(\pm) k}$ can also be found. Here we
propose the quantization scheme on the electromagnetic potentials
directly. One immediate merit is that the normalization condition of
the mode functions that will be used later to compute the number
density  of produced photons can be straightforwardly determined. It
is particularly useful to set up the initial-valued problem of the
nonequilibrium system where the initial mode functions can be
specified consistently subject to the  Wranskian conditions.
However, this huge simplification works only for the free photons.
Can neither $V^0_{1,2 k}$ nor $S_{(\pm)k}$ further diagonalize the
Lagrangian for the interactive  photon fields where the appropriate
perturbation expansion will be employed later.

As for the interactive photon fields under consideration, the
corresponding Lagrangian density  can be written in terms of the transverse degrees of freedom
as
\begin{equation}
L_{A}+L_{\zeta A}=  \frac{1}{2} \dot{A}_{T \, i} \dot{A}_{T}^i
-\frac{1}{2} \left(1- \frac{e^2 \lambda^2_V}{  m^2_{\pi} m^2_V}  \,
\dot\zeta^2  \right)\, \partial_l  A_{T\, i} \partial^l  A_{T}^i
-\frac{e^2}{4 \pi^2 f_{\pi}} \dot{\zeta} \frac{1}{\sqrt{g}}
\epsilon^{lmn} A_{T\, l} \, \partial_m  A_{T\, n} \, .
\end{equation}
>From observing the mode equations~(\ref{emv1}) and (\ref{emv2}), we
add and subtract the term
\begin{equation}
 L_{I} = \frac{1}{2} \, A_{T\,1 i} \left( \frac{2 \, \partial_{\bot}^2}{\tau^2  \partial^2 }-\frac{3 \, \partial^2_{\bot}}{ \tau^2 \partial^4}  \right)  A_{T\, 1}^i \, ,
\end{equation}
in the Lagrangian density $L_{A}+L_{\zeta A}$. The subtracted term
will modify the kinetic energy of the transverse modes of the vector
potentials with the polarization vector $\varepsilon^{(1)}_{{\bf
k}\,i}$ so that the expansion modes in terms of the circularly
polarized states to be defined later can diagonalize the
unperturbative Lagrangian density given by $ L_{A}+L_{\zeta A}-
L_{I}$  whereas the added term, $ L_{I}$,  is taken as the counter
term that will be treated perturbatively. The mode expansion of the
transverse vector potentials with the circularly polarized states
are given by
\begin{equation}
A_{T \, i} ({\bf{x}},\tau)= \int \frac{d^3 k}{( 2\pi)^{\frac{3}{2}}}
\frac{1}{\sqrt{2 k (\tau_0)}}  \left\{ \left[ b_{(+) {\bf k}}
V^0_{(+) k}(\tau) \varepsilon^{(+)}_{{\bf k} \,i}+ b_{(-) \bf k}
V^0_{(-) k}(\tau) \varepsilon^{(-)}_{ {\bf k} \,i} \right]
 e^{i\bf k\cdot \bf x} + {\rm h.c.} \right\}\, ,
\end{equation}
where the circular polarization vectors can be constructed out of
the linear polarization vectors $\varepsilon^{(1)}_{{\bf k} \,i}$
and $\varepsilon^{(2)}_{{\bf k} \,i}$ as follows:
\begin{equation}
\varepsilon^{(+)}_{{\bf k} \,i}=\frac{1}{\sqrt{2}} \left(i
\varepsilon^{(2)}_{{\bf k} \,i}+ \varepsilon^{(1)}_{{\bf k} \,i}
\right)\, , \,\,\, \varepsilon^{(-)}_{{\bf k}
\,i}=\frac{1}{\sqrt{2}} \left(i \varepsilon^{(2)}_{{\bf k} \,i}-
\varepsilon^{(1)}_{{\bf k} \,i} \right) \, .
\end{equation}
The corresponding mode equations obtained with respect to the
unperturbative Lagrangian density then become
\begin{eqnarray}
&& \ddot{V}^0_{(+) k} (\tau) +\frac{1}{\tau}  \dot{V}^0_{(+) k}
(\tau) +\left( 1-\frac{e^2 \lambda^2_V}{ m_{\pi}^2 m_V^2}
\dot\zeta^2 \right) k^2 V^0_{(+) k} (\tau)+ \frac{e^2}{ 2\pi^2
f_{\pi}} \,
\dot{\zeta} \, k \,  V^0_{(+) k} (\tau)=0 \, ,\label{emv+}\\
 && \ddot{V}^0_{(-)
k} (\tau) +\frac{1}{\tau}  \dot{V}^0_{(-) k} (\tau) +\left(
1-\frac{e^2 \lambda^2_V}{ m_{\pi}^2 m_V^2} \dot\zeta^2 \right) k^2
V^0_{(-) k} (\tau)- \frac{e^2}{ 2\pi^2 f_{\pi}} \, \dot{\zeta}\,  k
\, V^0_{(-) k} (\tau)=0 \, .\label{emv-}
\end{eqnarray}
Thus, as compared with the mode equations of free photons
(\ref{emv1}) and (\ref{emv2}), the involved interaction gives the
additional time dependent frequency terms in the mode equations that
depend on the dynamics of the expectation value of the neutral pion
field. In particular, when the mean field undergoes
the oscillations around the equilibrium value to be determined
dynamically, it is known that the solutions will exhibit the
features of the unstable bands and the growth of the fluctuating
modes. The growth of the modes in the unstable bands translates into
the profuse particle production. Thus, the photon production
mechanism is that of parametric amplification. Here the background
expansion will bring to us the extra damping effect. This is an
extension of our previous studies on parametric amplification of the
photon production to the case with the longitudinal background
expansion. A novel phenomenon can be observed. We also observe the
polarization asymmetry in the produced circularly polarized photons
as a result of the pseudo-scalar nature of the coupling that is of
interest to study.  It is conceivable that after the heavy ions
collision, the produced photons are mainly having large momentum
along the beam direction and then their $ k_\bot/k$ can be
perturbatively small. The perturbation expansion will be adopted for
this small parameter $ k^2_\bot/k^2$. The interaction Hamiltonain
density $ H_I$ can be constructed from the associated Lagrangian
density $L_I$ given by
\begin{eqnarray}
H_{I \, \bf{k}} (\tau) &=& \frac{1}{ 2 k (\tau_0)} \sqrt{g}
\frac{k^2_{\bot}}{ 2 k^2\tau^2}  \, \left[\,  b_{(+) \bf k}
{V}^0_{(+) k}(\tau) -
b_{(-) \bf k} {V}^0_{(-) k}(\tau) \right. \nonumber\\
&& \quad\quad\quad \left.+ b^{\dagger}_{(+) -\bf k} {V}^{0\,
*}_{(+) k}(\tau) -b^{\dagger}_{(-) -\bf k} {V}^{0\,
*}_{(-) k}(\tau)  \, \right] \, \times \,  \left[ \, \bf{k} \rightarrow
-\bf{k}\,  \right] \, .
\end{eqnarray}
Besides the interaction Hamilton depends on the
amount of  polarization asymmetry of the produced photons obtained
from the mode equations~(\ref{emv+}) and (\ref{emv-}) that certainly
give the next order results.

>From the exact expressions of the electric and magnetic fields in
Eqs.~(\ref{EField}) and (\ref{BField}), to order of $k^2_\bot/k^2$ they
can be obtained perturbatively as
\begin{eqnarray}
{{E}}^i_{\bf{k}} (\tau) &=&  {{E}}^i_{0 \, \bf{k}} (\tau) + \delta
{{E}}^i_{ \bf{k}} (\tau)+ 2 \,i \int d\tau'\,  \theta(\tau-\tau')\,
\left[ \,  {{E}}^i_{0 \,
\bf{k}} (\tau), \, H_{I \, \bf{k}} (\tau')\,  \right] +{\cal{O}} (k^4_\bot/k^4) \, ,\\
{{B}}^i_{\bf{k}} (\tau) &=&  {{B}}^i_{0 \, \bf{k}} (\tau) + 2\,  i
\int d\tau' \, \theta(\tau-\tau') \, \left[ \, {{B}}^i_{0 \, \bf{k}}
(\tau) ,\, H_{I \, \bf{k}} (\tau')\,  \right]  +{\cal{O}}
(k^4_\bot/k^4)\, ,
\end{eqnarray}
where
\begin{eqnarray}
E^i_{0\, \bf{k}} (\tau) &=& \frac{1}{\sqrt{2 k (\tau_0)}} \sqrt{g}
\left\{ \left[ \, \left( \, b_{(+) \bf k} \, \dot{V}^0_{(+) k}(\tau)
+ b^{\dagger}_{(+) -\bf k} \, \dot{V}^{0 \, *}_{(+)  k}(\tau)
\,\right)\, \varepsilon^{(+)\, i}_{\bf{ k} }\, \right]
+ \left[ \, (+) \rightarrow (-) \, \right] \right\} \,  , \label{E_0}\\
\delta E^i_{ \bf{k}} (\tau) &=& \frac{1}{\sqrt{2 k (\tau_0)}}
\sqrt{g} \frac{k^2_{\bot}}{ 2 k^2\tau^2}  \,  \left\{ \left[ \,
\left( \,b_{(+) \bf k} \, \dot{V}^0_{(+) k}(\tau) - b_{(-) \bf k} \,
\dot{V}^0_{(-) k}(\tau)
\right.\right.\right. \nonumber\\
&& \left.\left.\left. \quad\quad \quad\quad \quad\quad +
b^{\dagger}_{(+) - \bf k} \, \dot{V}^{ 0\,*}_{(+) k}(\tau)-
b^{\dagger}_{(-) - \bf k} \, \dot{V}^{0\, *}_{(-) k}(\tau) \right)
\varepsilon^{(+)\,
i}_{\bf{ k} } \, \right] + \left[\, (+) \rightarrow (-) \, \right] \right\} \,  ,\label{deltaE}\\
B^i_{0\, \bf{k}} (\tau) &=& \frac{-1}{\sqrt{2 k (\tau_0)}}
 \sqrt{g}  k  \left\{ \,
\left[\,\left(\,  b_{(+) \bf k} {V}^0_{(+) k}(\tau) +
b^{\dagger}_{(+) -\bf k} {V}^{0\, *}_{(+)  k}(\tau) \right)
\varepsilon^{(+)\, i}_{\bf{ k} } \,\right] - \left[ \, (+)
\rightarrow (-) \right]\, \right\} \, \label{B_0} \, .
\end{eqnarray}
Then, the backreaction effects from the photon
fields to the evolution of the mean field of the neutral pion in
Eq.~(\ref{eommf}) can be expressed by the electric and magnetic
fields that in turn can be obtained perturbatively with all above
results. They are
\begin{eqnarray}
&& \frac{1}{4} \,  \sqrt{g} \, \epsilon^{\alpha \beta \mu \nu}
\langle F_{\alpha \beta} F_{\mu \nu} \rangle (\tau) =  \langle \,
\left\{ \, E^i_{0 \, \bf{k}}(\tau) \, , \,    B_{ -{\bf{k}} \, i}
(\tau) \,\right\}\,
\rangle \nonumber \\
&&\quad\quad\quad\quad\quad\quad= \langle \,\left\{ \,  E^i_{0 \,
\bf{k}}(\tau) \,, \, B_{ -{\bf{k}} \, i} (\tau) \,\right\}\,\rangle
+ \langle \,\left\{ \, \delta E^i_{0 \, \bf{k}}(\tau) \,, \, B_{
-{\bf{k}} \,
i} (\tau) \,\right\}\,\rangle \nonumber\\
&&\quad \quad \quad \quad\quad\quad\quad+2\,  i \int d\tau'\,
\theta(\tau-\tau')\, \langle \, \left\{ \, E^i_{0 \, \bf{k}}(\tau)
\, ,  \left[ \, B_{0 \, {-\bf{k}}\, i} (\tau), \, H_{I \, {-\bf{k}}}
(\tau')\,  \right] \,\right\} \,
\rangle \nonumber \\
&&\quad\quad \quad\quad\quad\quad\quad\quad+2\,  i \int d\tau'\,
\theta(\tau-\tau')\, \langle \, \left\{ \, B^i_{0 \, \bf{k}}(\tau)
\, ,  \left[ \, E_{0 \, {-\bf{k}}\, i} (\tau), \, H_{I \, -\bf{k}}
(\tau')\,  \right]
\,\right\} \, \rangle \, +{\cal{O}} (k^4_\bot /k^4  ) \, , \nonumber \\
&& \frac{1}{4} \, \epsilon^{\mu\nu 0\sigma}\, \epsilon^{\alpha\beta
0\delta} \, g_{\sigma\delta}\,  \langle
F_{\mu\nu}F_{\alpha\beta}\rangle (\tau) =  \langle  \, \left\{ \,
B^i_{0 \, \bf{k}}(\tau) \, , \, B_{ -{\bf{k}} \, i} (\tau)
\,\right\}\, \rangle
\nonumber\\
&&\quad\quad = \langle \, B^i_{0 \, \bf{k}}(\tau) \, B_{0\,
-{\bf{k}} \, i} (\tau) \,\rangle +2\,  i \int d\tau'\,
\theta(\tau-\tau')\, \langle \, \left\{ \, B^i_{0 \, \bf{k}}(\tau)
\, ,  \left[ \, B_{0 \, {-\bf{k}}\, i} (\tau), \, H_{I \, {-\bf{k}}}
(\tau')\,  \right] \,\right\} +{\cal{O}} (k^4_\bot /k^4  ) \,
,\nonumber\\
&& \frac{1}{4} \, \epsilon^{\mu\nu 0\sigma}\,\epsilon^{\alpha\beta
\gamma\delta} \, g_{\sigma\delta} \, \langle
\partial_{\gamma} F_{\mu\nu}F_{\alpha\beta}\rangle (\tau)=\frac{d}{d\tau} \langle  \, \left\{ \, B^i_{0 \, \bf{k}}(\tau) \, , \, B_{
-{\bf{k}} \, i} (\tau) \,\right\}\, \rangle \, , \label{baphoton}
\end{eqnarray}
where the above correlation functions can be obtained  from
Eqs.~(\ref{E_0}), (\ref{deltaE}), and (\ref{B_0}), expressed in
terms of the mode functions, as follows:
\begin{eqnarray}
 &&\langle \,\left\{ \,  E^i_{0 \,
\bf{k}}(\tau) \,, \, B_{ -{\bf{k}} \, i} (\tau) \,\right\}\,\rangle
= \frac{1}{2 k (\tau_0) } \, g\, k \frac{d}{d\tau} \left[ \, \mid {V}^0_{(-) k}\mid^2 (\tau) -\mid {V}^0_{(+) k}\mid^2 (\tau)   \right] \,,\nonumber\\
 && \langle  \, \left\{ \, \delta E^i_{0 \, \bf{k}}(\tau) \, , \,     B_{
-{\bf{k}} \, i} (\tau) \,\right\}\, \rangle  =  \frac{1}{2 k
(\tau_0) } \, g \, \frac{k^2_{\bot}}{
   k \tau^2}  \, \left[\,  \mid {V}^0_{(-) k}\mid^2 (\tau)-   \mid {V}^0_{(+) k}\mid^2 (\tau) \,\right]\,,\nonumber\\
&&  \langle  \,\left\{ \,  B^i_{0 \, \bf{k}}(\tau) \,, \,  \left[ \,
E_{0 \, {-\bf{k}}\, i} (\tau) \,, \, H_{I \, -\bf{k}} (\tau')\,
\right] \, \right\}\, \rangle =  \frac{1}{2 k^2 (\tau_0) } \,
g^{\frac{3}{2}}\,
 \frac{k^2_{\bot}}{4 k \tau^2} \, \left\{\,\left[ \, \left( \, {V}^0_{(-) k}(\tau) {V}^{0 \, *}_{(-)  k}(\tau') +  \rm{c.} \, \rm{c.} \,\right) \right.\right.\nonumber\\
 &&\quad \quad\quad\quad\quad \quad\quad\quad \quad \quad\quad\quad\quad \quad\quad\quad\quad\quad\quad\left. \times \left( \, \dot{V}^0_{(-) k}(\tau) {V}^{0 \, *}_{(-)  k}(\tau') -  \rm{c.} \, \rm{c.} \,\right) \,-\, \left[\, (-) \rightarrow (+) \, \right]\,\right\}\,,\nonumber \\
&&  \langle  \,\left\{ \,  E^i_{0 \, \bf{k}}(\tau) \,, \,  \left[ \,
B_{0 \, {-\bf{k}}\, i} (\tau) \,, \, H_{I \, -\bf{k}} (\tau')\,
\right] \, \right\}\, \rangle =  \frac{1}{2 k^2 (\tau_0) } \,
g^{\frac{3}{2}}\,
 \frac{k^2_{\bot}}{4 k \tau^2} \, \left\{\,\left[ \, \left( \, \dot{V}^0_{(-) k}(\tau) {V}^{0 \, *}_{(-)  k}(\tau') +  \rm{c.} \, \rm{c.} \,\right) \right.\right.\nonumber\\
 &&\quad \quad\quad\quad\quad \quad\quad\quad \quad \quad\quad\quad\quad \quad\quad\quad\quad\quad\quad\left. \times \left( \, {V}^0_{(-) k}(\tau) {V}^{0 \, *}_{(-)  k}(\tau') -  \rm{c.} \, \rm{c.} \,\right) \,-\, \left[\, (-) \rightarrow (+) \, \right]\,\right\}\,,\nonumber \\
&&  \langle \, B^i_{0 \, \bf{k}}(\tau) \,\,  B_{0 \, -{\bf{k}} \, i}
(\tau) \,\rangle = \frac{1}{2 k (\tau_0)} \, g \, k^2 \,
\sum_{a=\pm}\, {V}^0_{(a) k}(\tau) {V}^{0 \,
*}_{(a)  k}(\tau) \,, \nonumber\\
&& \langle  \, \left\{ \, B^i_{0 \, \bf{k}}(\tau) \,,\,  \left[ \,
B_{0 \, {\bf{k}}\, i} (\tau)  \,,\,  H_{I \, \bf{k}} (\tau')\,
\right] \, \right\}
\, \rangle =  \frac{1}{2 k^2 (\tau_0)}\, g^{\frac{3}{2}} \,\frac{k^2_{\bot}}{4 \tau^2} \, \sum_{a=\pm}\, \, \left[ \, {V}^0_{(a) k}(\tau) {V}^{0 \, *}_{(a)  k}(\tau') +  \rm{c.} \, \rm{c.} \,\right] \nonumber\\
 &&\quad\quad \quad\quad\quad\quad\quad\quad \quad\quad\quad\quad\quad\quad \quad\quad\quad\quad\quad\quad \quad \quad\quad \times \left[ \, {V}^0_{(a) k}(\tau) {V}^{0 \, *}_{(a)  k}(\tau') -  \rm{c.} \, \rm{c.}
 \,\right]\,.
\label{correlationf}
\end{eqnarray}
The curly bracket in above means the anticommutator. The results
from the terms of order $k^2_\bot/k^2$ give an estimate on the
effects from their higher order contributions.

According to the Bjorken's scenario  the boost invariant
hydrodynamical flow might be created after the heavy ion
collisions. The corresponding initial vacuum state may already
contain particles with respect to their asymptotical states to be
observed in the detector~\cite{cooper}. Besides the relaxation of
the DCC can give an additional effect to produce the photons. Let
us now consider the initial state at time $\tau_0$ given by the
adiabatic modes in the comoving spacetime time that will be
specified below. The corresponding initial particle number density
can be expressed as
\begin{equation}
  \langle\, {\cal{N}}_k \, \rangle  (\tau_0) = \frac{1}{ 2 k (\tau_0) } \sqrt{g (\tau_0)} \left[ \, \langle \,
E^i_{\bf{k}}(\tau_0) \, E_{{-\bf k} \, i} (\tau_0) \, \rangle +
\langle \, B^i_{\bf{k}}(\tau_0) \, B_{{-\bf k} \, i} (\tau_0) \,
\rangle  \, \right] -1 \, .
\end{equation}
The expectation value of the number operator with respect to an
initial vacuum state  evolves in time and has the following form:
\begin{equation}
 \langle\, {\cal{N}}_k \, \rangle
(\tau)  = \frac{1}{ 2 k } \sqrt{g} \left[ \, \langle \, E^i_{{\bf
k}}(\tau) \, E_{-{\bf{k}} \, i} (\tau) \, \rangle + \langle \,
B^i_{\bf{k}}(\tau) \, B_{-{{\bf k}} \, i} (\tau) \, \rangle  \,
\right] -1 \,,
\end{equation}
that is, to order $k^2_\bot/k^2$,
\begin{eqnarray}
 \langle\, {\cal{N}}_k \, \rangle
(\tau)  &= & \langle \, E^i_{0 \, \bf{k}}(\tau) \, E_{0\, -{\bf{k}}
\, i} (\tau) \,\rangle +\langle \, B^i_{0 \, \bf{k}}(\tau) \, B_{0
\, -{\bf{k}} \, i} (\tau) \,\rangle+ \langle \,\left\{ \, E^i_{0 \,
\bf{k}}(\tau) \,, \,  \delta E_{ -{\bf{k}} \,
i} (\tau) \,\right\}\,\rangle \nonumber\\
&&\quad  +2\,  i \int d\tau'\, \theta(\tau-\tau')\, \langle \,
\left\{ \, E^i_{0 \, \bf{k}}(\tau) \, ,  \left[ \, E_{0 \,
{-\bf{k}}\, i} (\tau), \, H_{I \, {-\bf{k}}} (\tau')\,  \right]
\,\right\} \,
\rangle \nonumber \\
&&\quad\quad +2\,  i \int d\tau'\, \theta(\tau-\tau')\, \langle
\, \left\{ \, B^i_{0 \, \bf{k}}(\tau) \, ,  \left[ \, B_{0 \,
{-\bf{k}}\, i} (\tau), \, H_{I \, -\bf{k}} (\tau')\,  \right]
\,\right\} \, \rangle \, +{\cal{O}} (k^4_\bot /k^4  ) \, .
\label{nk}
\end{eqnarray}
Except for the correlation functions that we have found above, the
correlation functions between the electric fields are obtained as
\begin{eqnarray}
 && \langle \, E^i_{0 \, \bf{k}}(\tau) \,\, E_{0\, -{\bf{k}} \, i} (\tau)
\,\rangle
= \frac{1}{2 k (\tau_0) } \, g\, \sum_{a=\pm} \, \dot{V}^0_{(a) k}(\tau) \dot{V}^{0 \, *}_{(a)  k}(\tau) \,,\nonumber\\
 && \langle  \, \left\{ \, E^i_{0 \, \bf{k}}(\tau) \, , \,    \delta E_{
-{\bf{k}} \, i} (\tau) \,\right\}\, \rangle  =  \frac{1}{2 k
(\tau_0) } \, g \, \frac{k^2_{\bot}}{
  2 k^2 \tau^2}  \,\sum_{a=\pm}\, \frac{d}{d\tau} \mid {V}^0_{(a) k}\mid^2 (\tau) \,,\nonumber\\
&&  \langle  \,\left\{ \,  E^i_{0 \, \bf{k}}(\tau) \,, \,  \left[ \,
E_{0 \, {\bf{k}}\, i} (\tau) \,, \, H_{I \, \bf{k}} (\tau')\,
\right] \, \right\}\, \rangle =  \frac{1}{2 k^2 (\tau_0) } \,
g^{\frac{3}{2}}\,
 \frac{k^2_{\bot}}{4 k^2 \tau^2} \, \sum_{a=\pm}\, \, \left[ \, \dot{V}^0_{(a) k}(\tau) {V}^{0 \, *}_{(a)  k}(\tau') +  \rm{c.} \, \rm{c.} \,\right] \nonumber\\
 &&\quad \quad\quad\quad\quad\quad\quad \quad\quad \quad\quad\quad\quad \quad\quad\quad\quad\quad\quad \quad\quad \quad\quad\times \left[ \, \dot{V}^0_{(a) k}(\tau) {V}^{0 \, *}_{(a)  k}(\tau') -  \rm{c.} \, \rm{c.} \,\right] \,,
\label{correlationf}
\end{eqnarray}
 Thus the phase space number density in the comoving frame is given
by~\cite{boy2,leeng}
\begin{equation}
\frac{dN}{d\eta d^2 x_\bot d k_\eta d^2 k_\bot}  = \langle\,
{\cal{N}}_k \, \rangle \,.
   \label{conformalrate}
\end{equation}
We now need to relate this quantity to the invariant spectra of the
produced photons measured in the laboratory frame. From the
coordinate
 transformations~(\ref{transf}), we find that
\begin{eqnarray}
k_\tau &=& p_\bot \cosh( \eta-w)  \, , \nonumber \\ k_\eta &=& -
p_\bot \tau \sinh( \eta-w) \, , \nonumber \\  k_y &=& p_y ,
\nonumber\\
k_x &=& p_x ,  \label{eta-r}
\end{eqnarray}
where $k_\mu=(k ,{\bf k})$ is the photon four-momentum in the
comoving frame and $p_\mu=(p,{\bf p})$ is that measured in the
laboratory frame. We here also introduce the outgoing photon
particle rapidity defined by the four momentum in the center of the
mass coordinate system given by
\begin{equation}
p_\mu= ( \, p_\bot \cosh w \,,\,  {\bf p}_\bot \, , \,  p_\bot
\sinh w ) \, .
\end{equation}
Hence, from the momentum transformation laws, one can change the
momentum variables from $k_\mu$ into $p_\mu$ in the spectral number
density given by
\begin{equation}
p ~\frac{dN}{d^3 p }=\frac{dN}{d w d^2  k_\bot  }= \int dz \, d^2
x_\bot \, \Big{\vert} \frac{\partial k_\eta}{\partial
w}\frac{\partial \eta}{\partial z} \Big{\vert}\,  \frac{dN}{d\eta
d^2 x_\bot d k_\eta d^2 k_\bot }\, .
\end{equation}
Thus, carrying out the Jacobian, the end result of the invariant
spectra in the laboratory frame is obtained as~\cite{cooper}
\begin{equation}
p~ \frac{ dN}{d^3 p }=\frac{dN}{d w d^2  k_\bot  }= A_\bot \, \int
\, d k_\eta \,  \frac{dN}{d\eta d^2 x_\bot d k_\eta d^2 k_\bot }\,
,\label{spectra}
\end{equation}
where $ A_\bot$ is the transverse dimension of the effective
transverse size of the colliding ions. This quantity is
independent of $w$ as a consequence of the assumed boost
invariance.

Here we will postpone the full dynamical study on the photon
production to our future work and consider free photons instead.
With respect to an initial vacuum state of the photons given by the
adiabatic modes, the particle number can also be measured by their
asymptotical states to be observed in the detector under an
expanding background.  Exact analytical solutions to the mode
equations Eqs.(\ref{Spmk}) in the free photon case can allow to give
an order-of-magnitude estimate on this effect, and they also provide
an interesting example to illustrate the above formalism. To do so,
we express the expectation value of the number operator as follows:
\begin{eqnarray}
 \langle\, {\cal{N}}_k \, \rangle
(\tau) &=& \frac{\tau}{ 2 k (\tau_0) k} \left[\, \dot{S}_{(+) \, k}
(\tau) \dot{S}_{(+) \, k}^*  (\tau)+ \dot{S}_{(-) \, k}
(\tau)\dot{S}_{(-) \, k}^* (\tau) \right.\nonumber\\
&&\quad\quad\quad\quad\quad + \left. k^2 \left( {S}_{(+) \, k}
(\tau) {S}_{(+) \, k}^* (\tau)+ {S}_{(-)\, k}(\tau) {S}_{(-) \, k}^*
(\tau) \right) \right]-1\,.
\end{eqnarray}
The solutions $S_{(\pm)\, k)} $ to the equations~(\ref{Spmk}) can be
found analytically in below~\cite{sag}:
\begin{equation}
S_{(\pm)\,k} (\tau)= C_{(\pm) \, 1\,  k} \, e^{- \frac{\nu \pi}{2}}
H^{(1)}_{i\nu} (k_\bot \tau) + C_{(\pm ) \, 2 \, k} \, e^{-
\frac{\nu \pi}{2}} H^{(2)}_{i\nu} (k_\bot \tau) \, .
\end{equation}
Here $H^{(1)}_{i\nu}$ and $H^{(2)}_{i\nu}$ denote the Hankel
functions of the first and second kind with purely imaginary order,
$\nu=k_\eta$. Notice that the two solutions are complex conjugates
of each other. Thus, the above constants $ C_1$ and $C_2$ are to be
determined by the initial conditions. They can be obtained from
Eqs.~(\ref{SpmkV}) and (\ref{SdotpmkV}) in which the initial value
of the mode functions $V_{(1) \, k}$ and $V_{(2)\, k}$ are specified
by the adiabatic modes as~\cite{cooper}:
\begin{eqnarray}
V_{(1) \, k} (\tau_0) &=& \frac{e^{-i k(\tau_0)
\tau_0}}{\sqrt{\tau_0}} \,
, \,\,\,\,  \dot{V}_{(1) \, k} (\tau_0) =\frac{e^{-i k(\tau_0) \tau_0}}{\sqrt{\tau_0}} \left( -i k(\tau_0) -\frac{k_\bot^2}{ k^2(\tau_0) \tau_0} \right)\,, \\
V_{(2) \, k} (\tau_0) &=& \frac{e^{-i k(\tau_0)
\tau_0}}{\sqrt{\tau_0}} \,, \,\,\,\, \dot{V}_{(2) \, k} (\tau_0)
=\frac{e^{-i k(\tau_0) \tau_0}}{\sqrt{\tau_0}} \left( -i k(\tau_0)
\right) \,,
\end{eqnarray}
consistent with the Wranskian conditions~(\ref{wsv1}) and (\ref{wsv2}).
Fig.~1 displays the result of the spectra of the photons per unit
effective transverse area of the collisions associated with its
initial state measured with the reference of the asymptotical
states. Typically, less one particle  per unit
effective transverse area is found~\cite{lot}. It would be of
interest in both theoretically and experimentally to learn how the
photon production can be further amplified from the relaxation of a
DCC under a longitudinal background expansion.

\begin{figure}[htbp]
\begin{center}
\leavevmode  \epsfbox{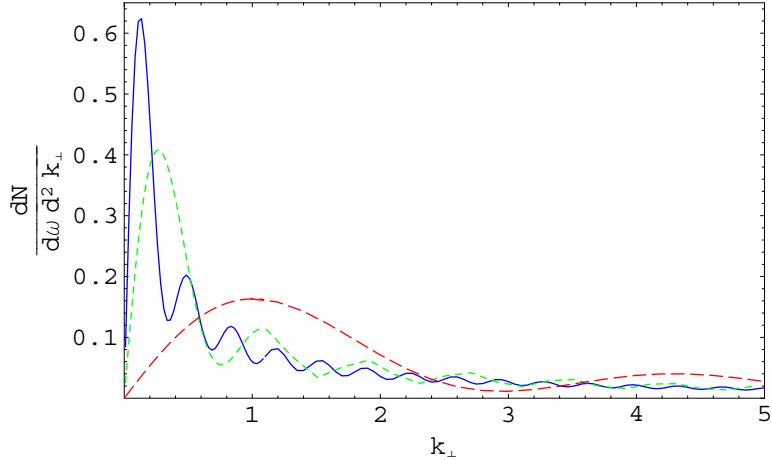}
\caption{ The respective spectra of the photons per unit effective
transverse area of the collisions  in its initial state measured by
the asymptotical states as a function of the transverse momentum (
in units of  $\tau_0$) at times ( in units of $\tau_0$ ) $\tau=2$ (
long-dashed), 5 ( short-dashed), 10 (solid).} \label{fig1}
\end{center}
\end{figure}

These derived dynamical equations display several important physical
processes: i) nonlinear relaxation of the mean field within a DCC,
ii) particle creation due to parametric amplification of the
fluctuations driven by the time dependent mean field, iii) the
enhancement from the boost effect under a longitudinal
hydrodynamical expansion. This must be studied self-consistently
from the coupled set of Eqs.~(\ref{eommf}), (\ref{pifluc}), and
(\ref{baphoton}) by inserting the mode functions in
Eqs.~(\ref{mepi}), (\ref{emv+}), and  (\ref{emv-}). The invariant
spectra of the observed photon production can be obtained from
Eqs.~(\ref{nk}), (\ref{correlationf}), and (\ref{spectra}).
 In particular, the expressions of corrections
  give an estimate on the ignored terms with higher order $k^2_\bot/k^2$.
   Although we do not intend to offer a numerical analysis of the
resulting equations in this paper, the nonequilibrium equation of
motion has the potential for providing an enhancement of the photon
production. This enhancement could lead to an experimentally
observable signal in the direct photon measurements of heavy-ion
collisions that can be a potential test of the formation of
disoriented chiral condensates. We postpone to a forthcoming paper
the full numerical study of these equations and an assessment of the
potential phenomenological impact of the nonequilibrium dynamics.

In conclusion, what we have done in this work is to propose the
quantization scheme on electromagnetic potentials under a
longitudinal background expansion, which suits the nonequilibrium
initial-valued problem. This is a necessary first step toward the
study of photon production through the non-equilibrium relaxation of
a disoriented chiral condensate formed in ultra-relativistic
heavy-ion collisions. The next calculation to do would be a
numerical study of the above equations with an eye on obtaining the
spectra of the produced photons that can be compared with thermal
photons from quark-gluon plasma and hadronic matter as done in
Ref.~\cite{charng}. This would have fascinating phenomenological
consequences.

\vskip 0.5cm

The work of D.-S. Lee and K.-W. Ng was supported in part by the
National Science Council, ROC under the grants
NSC97-2112-M-259-007-MY3 and NSC98-2112-M-001-009-MY3 respectively.

\end{document}